# Obtaining superhydrophobicity using commercial razor blades

Ken Yamamoto[1,2*], Hideyuki Takezawa[2], and Satoshi Ogata[2]

[1]Department of System Design Engineering, Keio University, 3-14-1 Hiyoshi, Kohoku-ku, Yokohama, Kanagawa 223-8522, Japan

[2]Department of Mechanical Engineering, Tokyo Metropolitan University, 1-1 Minami-Osawa, Hachioji, Tokyo 192-0397, Japan

*Corresponding author.
E-mail: yama-ken@tmu.ac.jp
Tel: +81 (0) 42 677 2710

**Abstract**

Because the superhydrophobic characteristic appears by forming a composite surface consisting of solid and air underneath the droplets, a large number of rough surfaces that can trap air have been fabricated. Recently, the air trapping on materials whose equilibrium contact angles are less than 90° was achieved by fabricating proper structures that lead energetic stability at the condition. Whereas these methods were proposed under the assumption of the static and equilibrium conditions, we take a dynamic and non-equilibrium approach in this study through droplet deposition and droplet impact experiments. By employing test surfaces that consist of commercially available stainless steel razor blades, we show the pinning effect brings the apparent water contact angle of approximately 160° on a "hydrophilic" substrate. We call this state the "non-equilibrium Cassie state" and give theoretical explanations. Furthermore, the dynamic characteristics of the droplet impact on these surfaces are discussed in a range of moderate Weber numbers ($We < 10^2$).

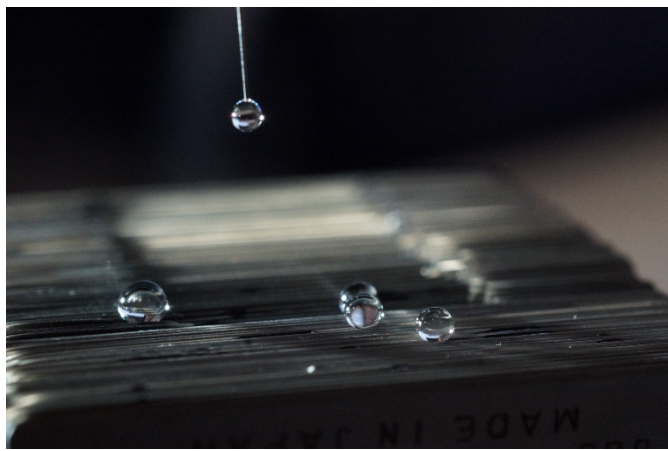

**INTRODUCTION**

Producing superhydrophobic surfaces, which show an apparent contact angle of more than 150°, have attracted a number of researchers for the past two decades [1]. Because the maximum limit of the equilibrium contact angle $\theta_{eq}$ on smooth surfaces consisting of a single material is 120° so far [2], superhydrophobicity can be achieved only by trapping air underneath the water droplet. This state, which is called the Cassie state [3], is generally obtained by roughening a surface made of a hydrophobic ($\theta_{eq} > 90°$) material [4, 5]. The wetting characteristic can also be changed by electrically or chemically coating the original substrate [6–9], and a combination of the surface roughening and the coating is often used to achieve the superhydrophobic surfaces [6, 9–13].

Furthermore, it is also possible to choose the Cassie state for surfaces having $\theta_{eq} < 90°$ by forming more complex topography [14–18]. Marmur [15] first defined theoretical criteria for the surfaces to take this state (often called the "metastable Cassie state" [16–18]) through the thermodynamic analysis and showed the state can be most stable in his model cases. However, despite the fact that $\theta_{eq}$ for metals is generally less than 90° (in other words they are "hydrophilic"), several "hydrophobic" metal surfaces with relatively simple topography and without coating were reported [19–22]. In contrast to the case of the metastable Cassie state, which is explained in the static (equilibrium) state, these studies include dynamic motions because droplets were released from some height. It implies the water repellency is also related to dynamic and non-equilibrium states.

Extrand [23] proposed a model for predicting the advancing contact angle $\theta_{adv}$ and receding contact angle $\theta_{rec}$ on structured surfaces that contains initial $\theta_{adv}$ and $\theta_{rec}$, surface geometry, droplet size, liquid density, and surface tension. A comparison of the predicted values with their experimental data showed that the model can accurately predict $\theta_{adv}$ and $\theta_{rec}$. Moreover, the model can predict the wetting state on structured surfaces. It suggests, even in the case of the static state, the dynamic contact angles have a significant role and the wetting transition has to be considered both statically and dynamically [17, 24, 25].

In this study, a droplet deposition experiment and a droplet impact experiment are performed for the sake of better understanding of the wetting transition. Superhydrophobic surfaces of $\theta_{app} > 150°$ are achieved with razor blade surfaces aligned in parallel and made of stainless steel, whose intrinsic contact angle is less than 90°, and the significance of the dynamic contact angles is discussed.

**EXPERIMENTAL SECTION**

The experiments were performed with two-dimensional structured stainless steel surfaces. The surfaces were prepared by manually aligning the commercial razor blades (FEATHER Safety Razor Co., Ltd., Hi-Stainless Platinum Double Edge Razor Blades) in parallel with some distance between the blades. Two surfaces were designed to have a constant spacing of ~100 μm and ~250 μm. We prepared one more surface that has random spacing of ~70–600 μm. The aligned blades were bonded by a commercial epoxy bonding agent. The surface roughness and blade angle of the razor blade were measured by a 3D laser scanning confocal microscope (KEYENCE, VK-X200) and determined as $R_a$ ~ 0.4 μm at the top of the blade, $R_a$ ~ 0.2 μm at the side plane of the blade, and angle of the blade (defined in Fig. 1) $2\alpha$ ~ 13°. The deviation of the blade tip positions in the direction of gravitational force (for descriptive purpose we call it "roughness") was also measured as $R_a$ ~ 25 μm and confirmed that the effect of the roughness was smaller than the measurement error in the advancing contact angle. Note that no chemical treatment was performed on the blades except for surface rinsing using ethanol and water. For almost all experiments, the initial diameter $D_0$ of the water droplet was set to 2.0 mm, which is smaller than the capillary length $L = (\sigma/\rho g)^{1/2}$ (2.7 mm for water droplets [26]), where $\rho$ and $\sigma$ denote the liquid density and the surface tension, respectively, and droplets of $D_0 = 2.7$ mm and 3.0 mm were employed in some cases for examining the effect of the droplet size and gravity. In the droplet deposition experiment, the produced droplets were released from a few hundred micrometers above the surfaces. All experiments were performed in the ambient air under controlled temperature of 20 ± 0.5°C. The relative humidity (RH) was ranged from 49.9% to 55.1% unless otherwise stated.

The intrinsic (equilibrium) contact angle and the advancing contact angle were measured on a flat stainless steel surface, which was also employed in the droplet impact experiment, because there are prints and coatings on the lateral side of the blades. The equilibrium contact angle was measured by recording droplets that were gently deposited on the surface, whereas the advancing contact angle was measured by recording the impact of droplets with the surface. The recordings were carried out using a high-speed camera (KEYENCE, VW-9000) and a microscope (KEYENCE, VH-Z35), with the help of a light source (HAYASHI, LA-HDF6010WD). Both angles were measured more than 10 times at different locations of the surface and then averaged.

The droplets were produced by flowing distilled water with a syringe pump (Chemyx Inc., NEXUS 6000) at a constant flow rate of 0.01 mL/min through one of three different needles. We used commercially available needles: flat-tipped 34G needles (ReactSystem) for droplets of $D_0 = 2.0$ mm, 21G and 18G needles (Terumo Japan) for $D_0 = 2.7$ mm and 3.0 mm, respectively. The resulting droplet volumes were 4 μL, 10.6 μL, and 15.2 μL (1.97 ± 0.035 mm, 2.72 ± 0.047 mm, and 3.07 ± 0.055 mm in diameter), respectively.

For droplet deposition experiment, water droplets with different diameters were dropped on test surfaces from a few hundred micrometers above the surfaces. Under this condition, the droplets never bounded. The deposition of the droplet was recorded with the same setting as the one used to measure the contact angle. On the other hand, for droplet impact experiment, water droplets with different diameters were dropped on test surfaces from ~1–100 mm above the surfaces. The impact of the droplet was recorded with the same setting used to measure the contact angle. The images were recorded at a frame rate of 4000–15000 fps with a magnification of 50×–200×.

**RESULTS AND DISCUSSION**

As seen in Fig. 2, the razor blade surfaces showed superhydrophobicity with the apparent contact angle $\theta_{app}$ of 147–171° and a significantly small contact angle hysteresis, whereas the equilibrium contact angle $\theta_{eq}$ of the stainless steel was less than 67°. The measured $\theta_{app}$ was roughly similar to the estimated value of 160.7°, which was calculated from the Cassie-Baxter equation: $\cos\theta_{app} = f(\cos\theta_{eq} + 1) - 1$ [3], where $f$ denotes the area fraction of the solid (estimated to be ~4%). Because the equation expresses the heterogeneity of a flat plane, the correspondence means that the three-phase (water–air–solid) contact lines and water–air interfaces on the bottom of the droplet are at the same level as the position of the blade tips. Indeed, we observed that the three-phase contact lines were pinned at the edge of the blades, with angles larger than $\theta_{eq}$. Therefore, we assume that pinning occurs when the advancing contact angle $\theta_{adv}$ is larger than the threshold angle $\theta_{pin} = \alpha + 90°$, which is determined by the edge angle as illustrated in Fig. 3a. With this criterion we can predict whether the droplet has the possibility to take the Cassie state on a surface. Because this state is not the conventional Cassie state nor "metastable" Cassie state, and because this state is energetically non-equilibrium, in contrast to former two states, we call it "non-equilibrium Cassie state". Moreover, the blade spacing $w$ has to be considered in the case that $D_0$ and $w$ are comparable. In this case, the radius of the water–air interface between the blades $R_1$ can be geometrically derived as

$$R_1 = \frac{w}{2\sin(\theta_{adv} - 90° - \alpha)} = -\frac{w}{2\cos(\theta_{adv} - \alpha)}, \qquad (1)$$

and the droplet cannot stay on the edges if $R_1$ is larger than the initial droplet radius $R_0$ (Fig. 3b). Therefore, the maximum $w$ can be obtained as $w_{max} = -2R_0\cos(\theta_{adv} - \alpha)$, or in a non-dimensional form

$$w_{max}/D_0 = -\cos(\theta_{adv} - \alpha). \qquad (2)$$

Employing Eq. (2), we obtained the predicted value $w_{max}/D_0 = 0.42$ with $\theta_{adv} = 121.4°$ and $\alpha = 6.5°$. In most cases, the measured $w_{max}/D_0$ was similar to the predicted value and in some cases, it exceeded 0.42 (the maximum value was 0.52) due to the uncertainty of the advancing contact angle ($\theta_{adv} = 121.4 \pm 6.3°$). By considering this deviation in $\theta_{adv}$, we obtain $w_{max}/D_0 = 0.50$ as the upper limit, which is close to the experimental observation.

From the above discussion, we derived a two-step criterion for droplets on two-dimensional structured surfaces to take the non-equilibrium Cassie state as (a) $\theta_{adv} > \theta_{pin} = \alpha + 90°$ and (b) $w < -D_0\cos(\theta_{adv} - \alpha)$. This criterion indicates that a combination of a sharp edge angle and narrow blade spacing is effective for the droplets to remain on the edges.

Subsequently, a droplet impact experiment was performed in order to understand the wetting transition from a dynamic approach. In this experiment, the initial height of the droplet to be released was varied, thereby various droplet velocities at the collision with the surfaces $U$ were examined. A typical sequence of the droplet impact on the surface is shown in Fig. 2 (left column). We observed that the droplets bounced several times when the Weber number $We$ (= $\rho U^2 D_0/\sigma$) was less than five.

For $We < 5$, the three-phase contact lines were completely pinned on the edges of the blades (Fig. 2, left column). Furthermore, the behavior of the advancing interface (Fig. 2, center-left column) agreed with the model proposed by Extrand [23]. This indicates the validity of the proposed criterion based on the droplet deposition experiment because the model was derived while considering $\theta_{adv}$. On the other hand, the receding interface (Fig. 2, center-right column) did not agree with the model because of the volume of the droplet that limited the practical receding angle. However, the experimental results agreed with the experimental observations examined in quasi-static conditions [27, 28].

For $We > 5$, blade spacing of relatively wide $w$ was not able to pin the three-phase contact line on the edges and water penetrated into the grooves. For higher $We$, almost all blade spacing could not hold the water–air interface at the edges (Fig. 2, right column) and some droplets divided into multiple droplets as it was reported earlier [29, 30]. Figure 4 shows a relationship between $We$ and $w/D_0$. We found from Fig. 4 that there is a threshold value of $w/D_0$ that depends on $We$. Because $We$ indicates the balance of inertia and surface tension, we equated the dynamic pressure $P_d = \rho U^2/2$ [31] and the difference between the Laplace pressures at the top and bottom of the droplet calculated from $R_0$ and $R_1$ as

$$\frac{1}{2}\rho U^2 \approx \frac{2\sigma}{R_1} - \frac{2\sigma}{R_0}, \tag{3}$$

and we obtained the threshold by substituting Eq. (1) into Eq. (3),

$$\frac{w}{D_0} \approx \frac{-4\cos(\theta_{adv} - \alpha)}{We + 4}. \tag{4}$$

The threshold calculated by Eq. (4), represented as a dashed line in Fig. 4, shows good agreement with the experimental result, and it corresponds to Eq. (2) when $We = 0$. Moreover, the same experiment with $D_0 = 3.0$ mm suggests that the relationship still held even when $D_0$ was larger than the capillary length. We also performed the same experiment on a rough copper surface ($\theta_{adv} = 114°$) prepared by the wire-EDM (Fig. 5) for verifying that the proposed model is not a particular-product-dependent. The surface has two-dimensional saw-tooth roughness with the edge angle (equivalent to the blade angle) $2\alpha = 26°$ and the edge spacing of 500 μm. Although we could examine with only two droplet diameter ($D_0 = 2.0$ mm and 2.7 mm) due to the limitation of the fabricated area (3.0 mm). However, because the threshold line derived from Eq. 4 was found to be valid on the surface as well (Fig. 5c), thereby suggesting that the proposed model does not depend a specific product.

In the discussions so far, we showed how easily superhydrophobicity is achievable and how important the advancing contact angle is in the static condition. Now we will discuss the effect of the hydrophilicity of the substrate on the behavior of the droplets in a dynamic condition. Liu et al. [32] fabricated pillar-forested surfaces, of which pillars are hydrophobic and located a few hundred micrometers away from each other, and conducted an experiment similar to our droplet impact experiment, but with higher $We$. As a result, they observed that water penetrated into the pillar forest at $We \sim 10$ and a recoil of the penetrated water in the vertical direction occurred afterwards owing to the pillar's widening structure. This is essentially impossible for our surfaces because of their hydrophilic characteristic on the lateral sides of the blades. Although this characteristic and the droplet break-up [29, 30] limit a range of $We$ to show the hydrophobic characteristic for the razor blade surfaces up to $\sim 10^0$, it may work for the top edges of the surfaces to keep dry by the complete penetration of water between the blades instead of the transition to the Wenzel state on supherhydrophobic surfaces [12, 24, 25].

Figure 6 shows relationships between $We$ and the time droplet contacting the surface $T_c$ and the restitution coefficient $\varepsilon$, which is defined as $\varepsilon = U/U_b$, where $U_b$ denotes the velocity after the droplet bounded [29]. As it was observed on structured hydrophobic surfaces [26, 32, 33], $T_c$ decreases in a region where $We < 1$ and it takes constant value, which is scaled as $(\rho R_0^3/\sigma)^{1/2}$ at $We > 1$ (Fig. 6a). We confirmed the same tendency with droplets of diameters $D_0 = 2.7$ mm and 3.0 mm as well. A transition at $We \sim 1$ is also observed in the relationship between $We$ and $\varepsilon$ (Fig. 6b). It can be seen that $\varepsilon \sim 0.9$ below $We \sim 1$, and it decreases with an increase in $We$, which was also observed on the structured hydrophobic surfaces [29, 34]. Note that $\varepsilon$ was calculated as $(h/h_b)^{1/2}$ in our experiment, where $h$ and $h_b$ denote the initial height the droplet is released from and the maximum height of the bounced droplet, respectively, because the measurement of $U_b$ was difficult owing to the excessive deformation of the droplet. These results suggest that the kinetic energy of the impact is stored in the surface energy as is the case with the hydrophobic surfaces.

As a droplet hits a surface, it deforms into a pancake-like shape. By measuring the maximum diameter of the pancake $D_{max}$, we explored a dependency of $D_{max}$ (in a normalization form of $D_{max}/D_0$) on $We$ (Fig. 7). We also performed the same experiment on a smooth and flat stainless steel surface and the result is represented as a reference in Fig. 7. It is observed from Fig. 7 that $D_{max}/D_0$ on the flat surface shows the relationship $D_{max}/D_0 \sim We^{1/4}$ for $We > 10$ as Claret et al. [33] derived by scaling analysis, although the surface was hydrophilic and the droplet on it did not detach from the surface after the impact. On the other hand, for smaller $We$, droplets on both the razor blades and the flat surface follow $We^{1/2}$ lines, implying a large part of the initial kinetic energy was stored in the surface energy [26, 33]. This conclusion corresponds to the conclusion drawn from Fig. 6. Moreover, the reason that the horizontal deformation of the droplets on the razor blades are significantly smaller than that on the flat surface can be explained as follows: the correlation $D_{max}/D_0 \sim We^{1/2}$ was derived by equating the kinetic energy (scaled as $\rho U^2/R_0$) and the gradient of the Laplace pressure (scaled as $\sigma/H$, where $H$ denotes the height of the pancake) with a correlation $R_0^3 \sim HR_{max}^2$, derived from the volume conservation [26]. Therefore, the small horizontal deformation suggests a larger deformation in vertical direction. In fact, we observed large deformations of the interfaces pinned on the blade edges (Fig. 8). From these observations, we conclude that most of the kinetic energy from the impact is stored in the surface energy when $We < 10$ and RH $< 55\%$, and it is possible to produce stable and irrefrangible superhydrophobic surfaces with metals that are comparable with the surfaces made of hydrophobic substrates.

However, the characteristics were changed under a condition of high RH (67.78 ± 2.76%). In this case the penetration by the droplets occurred with smaller $w/D_0$ than the threshold for all range of $We$ (Fig. 9a), while the advancing contact angle remained almost the same (118.9 ± 3.2°). Moreover, the droplets on the razor blade surfaces start penetrating at $We \sim 5$ ($We \sim 15$ for low RH) and the transition of $D_{max}/D_0$ from $We^{1/2}$ line to $We^{1/4}$ line was observed for smaller $We$ (Fig. 9b). Because this transition suggests the amount of the viscous dissipation became relatively large, we compared a duration from the impact to the moment that the pancake radius reaches its maximum (expressed as $T_{0-max}$) on the flat surface for high (68%) and low (50%) RH (Fig. 10). Consequently, it is observed that $T_{0-max}$ for high RH is approximately 85% of that for low RH. It suggests that the velocity of the three-phase contact line increases as RH increases and the viscous dissipation for high RH is approximately 15% higher than that for low RH. From these observations we concluded that this larger viscous dissipation induced early transition to $We^{1/4}$ line. Furthermore, it is implied that the pinning force is a RH-dependent because the three-phase contact line can advance faster as if the wall friction is decreased as RH increases.

**CONCLUSIONS**

We fabricated the superhydrophobic surfaces with commercially-available razor blades made of stainless steel that show an apparent contact angle of 147–171°. Because this superhydrophobic state is achieved due to the advancing (i.e., non-equilibrium) contact angle, it was named the non-equilibrium superhydrophobic state. Criteria to obtain the state were theoretically derived based on the surface geometry and the advancing contact angle. The dynamic characteristics of the droplets on the surfaces were investigated. It was revealed that whether the three-phase contact line can remain pinned is determined by the balance of the dynamic pressure and the amount of the difference in the Laplace pressures at the top and bottom of the droplet. We confirmed through the droplet impact experiment that the behavior of the droplet on the fabricated surfaces is similar to that on surfaces made of hydrophobic substrates as long as the droplet remained united and pinned on the surface edges. Although this aspect limits the applicability of the surfaces to *We* < 10, they are comparable to the hydrophobic-material-based surfaces because of their high stiffness and their switchable wettability. Furthermore, because the achieved superhydrophobicity can only exist in the non-equilibrium state, effects of the relative humidity (RH) was examined. Consequently, it was concluded from the droplet impact experiment on the flat surface that the pinning force weakened under high RH while the advancing contact angle remained the same as that in moderate RH condition.

**ACKNOWLEDGEMENTS**

This work was supported by JSPS KAKENHI Grant Number 22760134 and TMU Young Scientist Research Fund (2014). We also would like to acknowledge Mr. Kozaburo Ishii, the president of the ReactSystem, for providing custom-made flat tip needles.

FIGURES

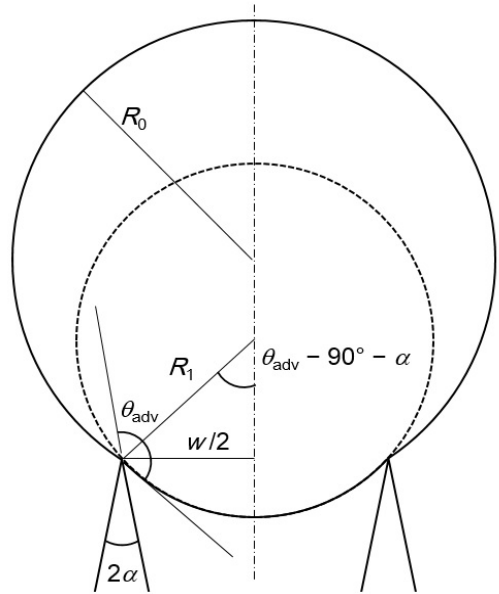

Figure 1. A schematic model for the criterion, to verify if the interface can be pinned. The interface can be pinned when the apparent contact angle $\theta_a$ is larger than $\theta_{pin}$.

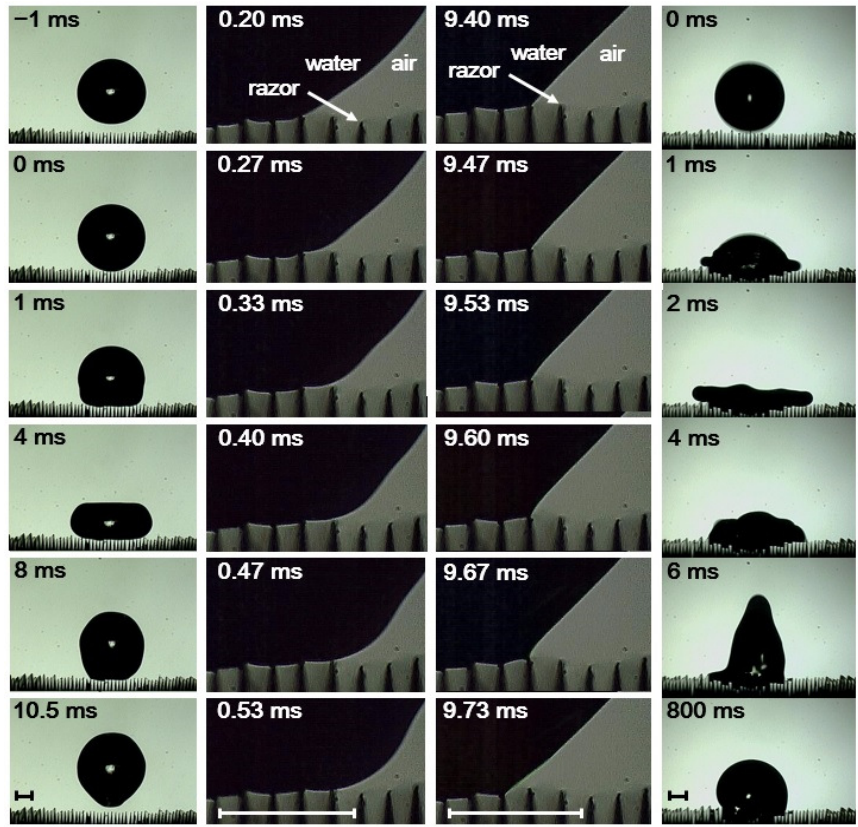

Figure 2. Droplet impact on the razor surface (scale bars represent 500 µm). Left column: successive images of a bouncing droplet ($D_0 = 2.0$ mm, $We = 1.34$) recorded by a high-speed camera at 4000 fps and with a magnification of 50×. Center-left column: successive images of an advancing interface of a droplet in the same condition captured at 15000 fps and with a magnification of 200×. Center-right column: successive images of a receding interface of the same droplet captured at 15000 fps and with a magnification of 200×. Right column: successive images of a penetrating droplet ($D_0 = 2.0$ mm, $We = 27.60$) impacting the same location as the left column droplet (recorded at 4000 fps and 50×).

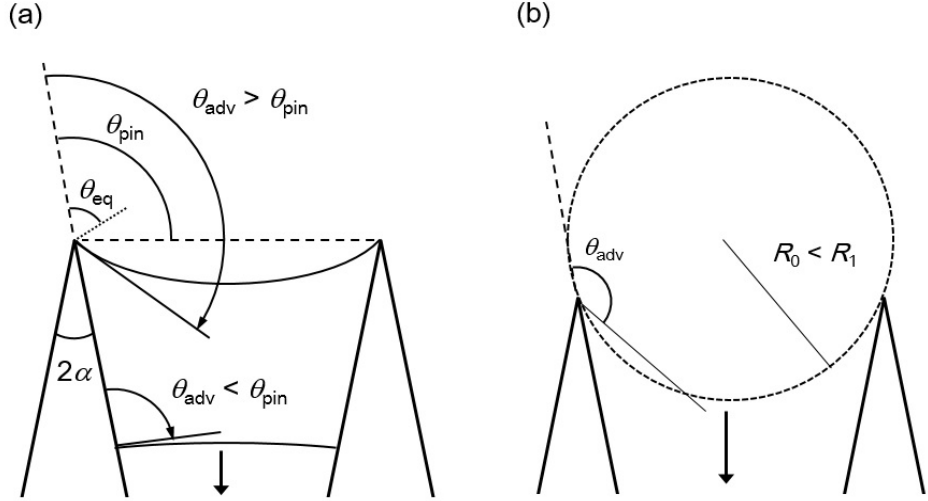

Figure 3. (a) A schematic model for a large deformation of the interface between the edges. The pinned interface can largely deform when the spacing between the edges $w$ is comparable to the droplet diameter. An assumption that the contact angle at the edge is equal to $\theta_a$ gives the minimum curvature of $1/R_1$. (b) A schematic model for the criterion of the maximum width that can hold the droplet (in case of $D_0 < L$). The droplet will penetrate the lateral surface of the blades when $R_0 < R_1$.

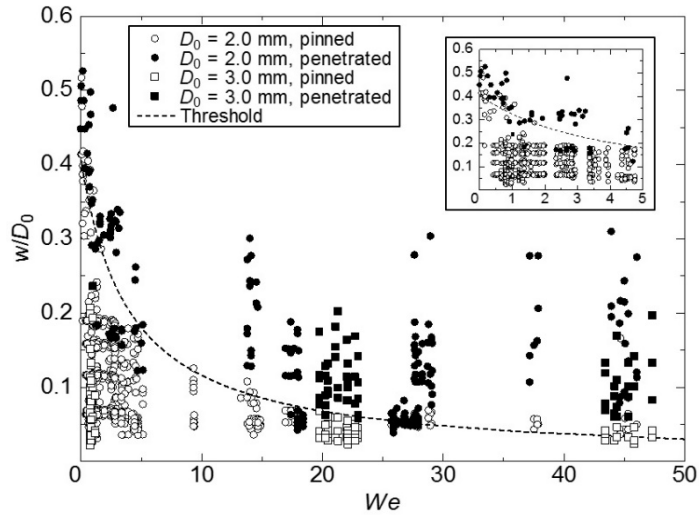

Figure 4. Relationship between $We$ and $w/D_0$. The interface of the impacting droplet can be pinned if the threshold $w/D_0 < -4\cos(\theta_{adv} - \alpha)/(We + 4)$ (shown as a dashed line). The experiment was performed with three different droplet diameters ($D_0$ = 2.0 mm, 2.7 mm, and 3.0 mm) and it was confirmed that the threshold was valid for all the cases (only results with $D_0$ = 2.0 mm and 3.0 mm are shown in the figure for a better visibility). The inset shows a zoom-up of the region where $We$ = 0–5.

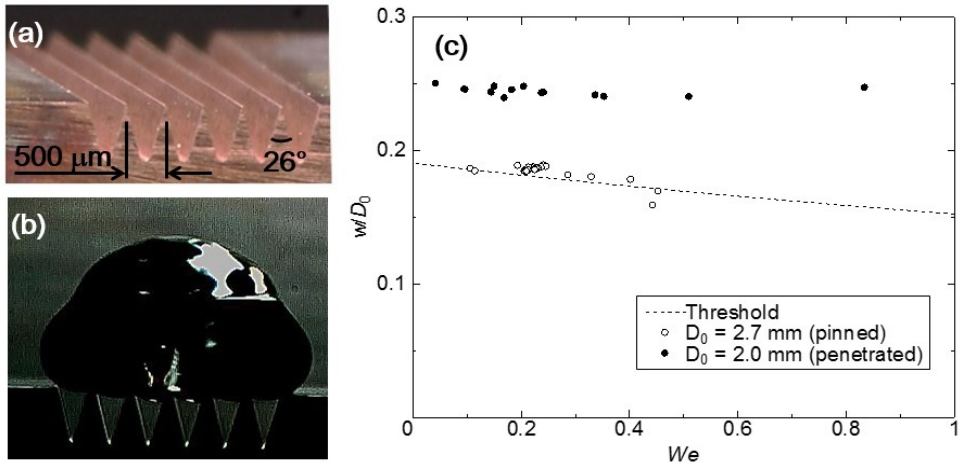

Figure 5. Droplet impact experiment with a copper saw-tooth surface. (a) Structure of the surface. The surface was fabricated by the wire EDM. The edge distance and angle are 500 μm and 26°. (b) A snapshot of the bouncing droplet ($D_0$ = 2.7 mm). (c) The relationship between $We$ and $w/D_0$. The value of $w/D_0$ was varied by changing the droplet size ($D_0$ = 2.0 mm or 2.7 mm). The dashed line indicates the threshold line calculated from Eq. 4 with $\theta_{adv}$ = 114° and $\alpha$ = 13°. The result shows a good predictability of the proposed model even on the copper saw-tooth surfaces.

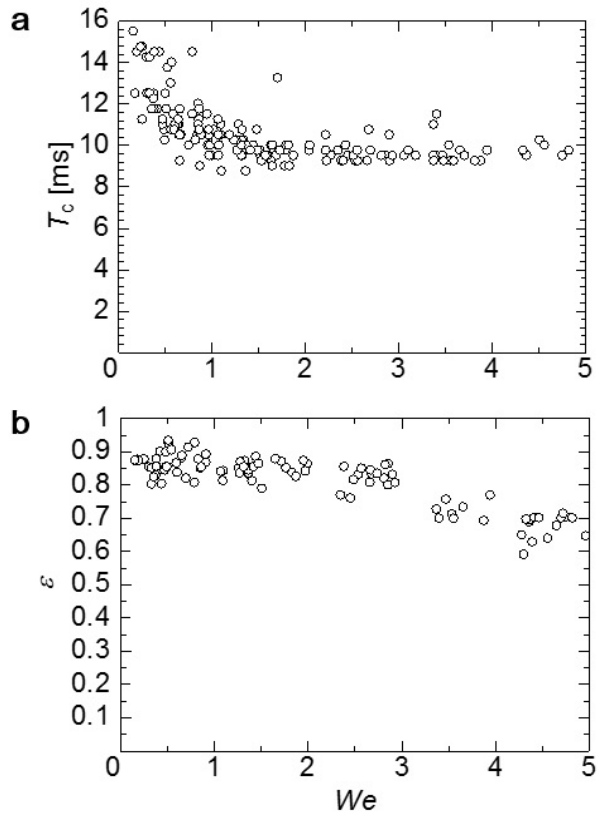

Figure 6. Dependency of the dynamic characteristics of the bouncing droplets on $We$. (a) A relationship between *We* and the contacting time $T_c$. $T_c$ shows a dependency on *We* in the region where *We* < 1, while it is Weber-number-independent in the region where *We* > 1. (b) Relationship between *We* and the restitution coefficient $\varepsilon$. The coefficient shows ~0.9 in the region where *We* < 1 and it decreases with an increase in *We*.

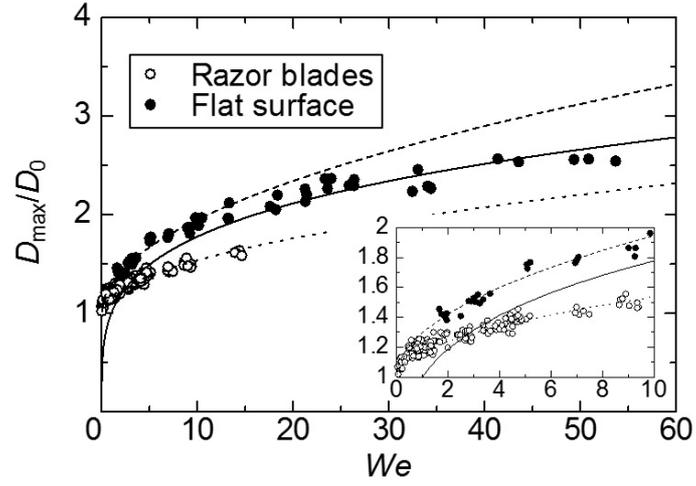

Figure 7. Dependency of the maximum droplet diameter $D_{max}$ on *We* (in a form of $D_{max}/D_0$). Open and closed circles indicate droplets on the razor blades and on the stainless flat surface, respectively. The solid line represents $We^{1/4}$ ($D_{max}/D_0 = We^{1/4}$), and the dashed and dotted lines represent $We^{1/2}$ ($D_{max}/D_0 = 1 + 0.3We^{1/2}$ and $D_{max}/D_0 = 1 + 0.17We^{1/2}$), respectively. The inset represents a zoom-up of the region where *We* = 0–10.

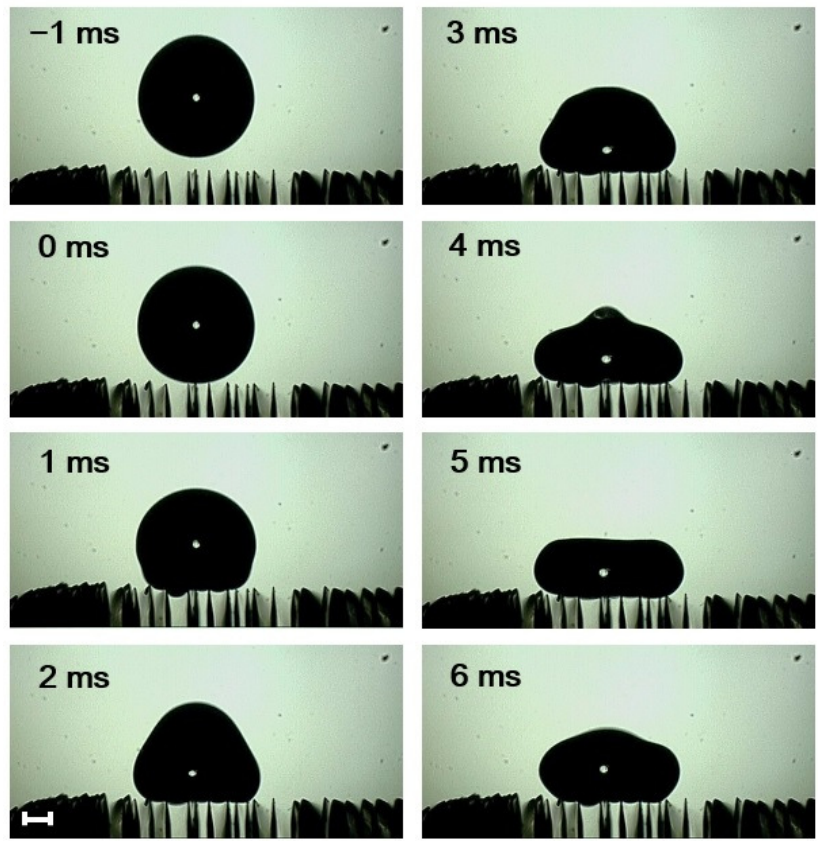

Figure 8. Oscillations of the pinned interfaces of a bouncing droplet on a randomly spaced blade surface. The frequency of the oscillation is higher (~4 times) than that of the droplet bounce. Scale bar represents 500 μm.

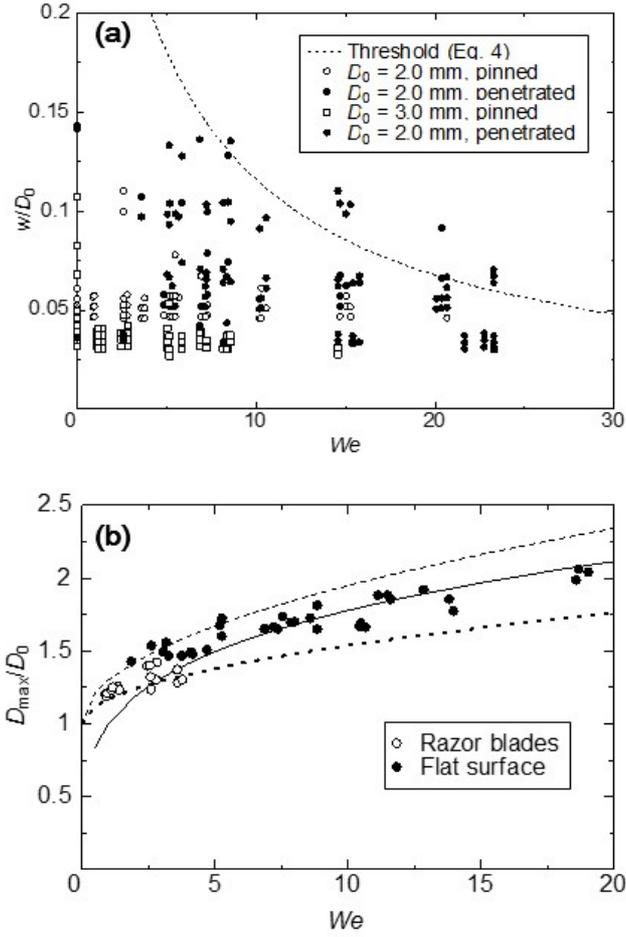

Figure 9. Droplet impact experiment at humid condition (RH = 67.8 ± 2.76%). (a) The relationship between $We$ and $w/D_0$. Although $\theta_{adv}$ (= 118.9 ± 3.2°) was hardly changed from low RH condition (= 121.4 ± 6.3° at RH ~ 50%), the droplets penetrate narrower spacing than predicted threshold (Eq. 4). (b) The relationship between $We$ and $D_{max}/D_0$. The solid line represents $We^{1/4}$ ($D_{max}/D_0 = We^{1/4}$), and the dashed and dotted lines represent $We^{1/2}$ ($D_{max}/D_0 = 1 + 0.3We^{1/2}$ and $D_{max}/D_0 = 1 + 0.17We^{1/2}$), respectively. The droplets impacting the flat surface show faster transition from $We^{1/2}$ line to $We^{1/4}$ line than in the case of low RH (Fig. 7).

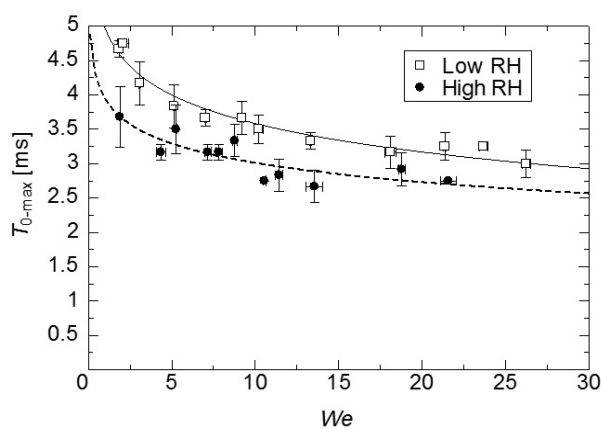

Figure 10. Effect of RH on the duration of the three-phase contact line advancement $T_{0-max}$ (from impact to the moment that horizontal droplet diameter becomes $D_{max}$). Solid and dashed lines are fitted curves for Low RH (~ 50%) and High RH (~ 68%).